\documentclass[twocolumn,showpacs,preprintnumbers,amssymb]{revtex4-1}
\usepackage{graphicx}
\usepackage{dcolumn}
\usepackage{color}
\usepackage{bm}

\def \a{\alpha}
\def \b{\beta}
\def \l{\lambda}
\def \L{\Lambda}

\def \s{\sqrt}
\def \be{\begin{equation}}
\def \ee{\end{equation}}
\def \ben{\begin{eqnarray}}
\def \een{\end{eqnarray}}
\def \o{\omega}

\def \D{\Delta}
\def \S{\Sigma}

\begin{document}

\title{The Hawking temperature in the context of dark energy for Kerr-Newman and Kerr-Newman-AdS backgrounds}

\author{Goutam Manna }
\altaffiliation{goutammanna.pkc@gmail.com}
\affiliation{Department of Physics, Prabhat Kumar College, Contai, Purba Medinipur-721404, India}

\author{Bivash Majumder}
\altaffiliation{bivashmajumder@gmail.com}
\affiliation{Department of Mathematics, Prabhat Kumar College, Contai, Purba Medinipur-721404, India}

\begin{abstract}
 We show that the Hawking temperature is modified in the presence of dark energy in an emergent gravity scenario for Kerr-Newman(KN) and Kerr-Newman-AdS(KNAdS) background metrics. 
The emergent gravity metric is not conformally equivalent to the gravitational metric. We calculate the Hawking temperatures for these emergent gravity metrics along $\theta=0$. Also we show that the emergent black hole metrics are satisfying Einstein's equations for large $r$ and $\theta=0$. Our analysis is done in the 
context of dark energy in an emergent gravity scenario having $k-$essence scalar fields $\phi$ with a Dirac-Born-Infeld type lagrangian. In KN and KNAdS background, the scalar field $\phi(r,t)=\phi_{1}(r)+\phi_{2}(t)$  satisfies the emergent gravity equations of motion at  
$r\rightarrow\infty$ for $\theta=0$.



\end{abstract}
\maketitle

\section{Introduction}
Research on the context of the Hawking temperature has gained momentum during last two decades.
It has been shown that the Hawking temperature \cite{haw} is modified in the presence of dark energy in an emergent gravity scenario for Schwarzschild, Reissner-Nordstrom and Kerr background in \cite{gm1,gm2}. As seen in \cite{gm1,gm2}, for an emergent gravity metric $\tilde G_{\mu\nu}$ is conformally transformed into $\bar G_{\mu\nu}$ where $\bar G_{\mu\nu}= g_{\mu\nu} - \partial _{\mu}\phi\partial_{\nu}\phi$ ($g_{\mu\nu}$ is the gravitational metric) for Dirac-Born-Infeld(DBI) \cite{born} type Lagrangian having $\phi$ as $k-$essence scalar field. The Lagrangian for $k-$essence scalar fields contains non-canonical kinetic terms. 
The general form of the Lagrangian for $k-$essence model is: $L=-V(\phi)F(X)$ where $X=\frac{1}{2}g^{\mu\nu}\nabla_{\mu}\phi\nabla_{\nu}\phi$ and it does not 
depend explicitly on $\phi$ to start with \cite{gm1,gm2,babi,scherrer}. 

Relativistic field theories with canonical kinetic terms have the distinction from those with non-canonical kinetic terms associated with $k-$essence, since the nontrivial dynamical solutions of the k-essence equation of motion not only spontaneously break Lorentz invariance but also change the metric for the perturbations around these solutions. Thus the perturbations propagate in the so called {\it emergent} or analogue curved spacetime \cite{babi} with the metric different from the gravitational one. Relevant literatures  \cite{gorini} for such fields discuss about cosmology, inflation, dark matter, dark energy and strings. 

The motivation of this work is to calculate the Hawking temperature in the presence of dark energy for an emergent gravity metric which is also a blackhole metric. We consider two cases, (a) when the gravitational metric is a Kerr-Newman and (b) when the gravitational metric Kerr-Newman-AdS.

In \cite{umetsu}-\cite{aliev}, discuss about Hawking radiation for Kerr, Kerr-Newman, Kerr-Newman-AdS etc. black holes using different techniques. Here we calculate the Hawking temperature for emergent gravity metric for Kerr-Newman and Kerr-Newman-AdS backgrounds using tunneling mechanism. These temperatures are different from usual temperatures of Kerr-Newman and Kerr-Newman-AdS black holes.

In section 2, we have described $k-$essence and emergent gravity where the metric $\tilde G_{\mu\nu}$ contains the 
dark energy field $\phi$ and this field should satisfy the emergent gravity equations of motion. 
Again, for  $\tilde G_{\mu\nu}$ is to be a blackhole metric, it has to satisfy the Einstein field equations. The formalism for $k-$essence and emergent gravity used is as described in \cite{babi}.

In section 3 and 5, we have shown that for Kerr-Newman and Kerr-Newman-AdS both cases, the emergent gravity metrics are mapped on to the Kerr-Newman and Kerr-Newman-AdS type metrics in the presence of dark energy. The emergent metric  satisfies Einstein equations for large $r$ and the  dark energy field $\phi$ satisfies the emergent gravity equations of motion for along $\theta=0$ at $r\rightarrow\infty$.  

We have calculated the Hawking temperature for emergent gravity metrics for Kerr-Newman and Kerr-Newman-AdS backgrounds in section 4 and 6 respectively. We clarify that the Hawking temperature is spherically symmetric from very general conditions and taking $\theta=0$ does not therefore affect this property of the Hawking temperature.  It has been shown elaborately in \cite{mann}, how the Hawking temperature is independent of $\theta$, although the metric functions depend on $\theta$. Also Hawking temperature is purely horizon phenomenon of the spacetime where the temperature is not depending on $\theta$. So we can say that the Hawking temperature is spherically symmetric.

\section{$k-$essence and Emergent Gravity}
The $k-$essence scalar field $\phi$ minimally coupled to the gravitational field $g_{\mu\nu}$ has action \cite{babi}
\ben
S_{k}[\phi,g_{\mu\nu}]= \int d^{4}x {\sqrt -g} L(X,\phi)
\label{eq:1}
\een
where $X={1\over 2}g^{\mu\nu}\nabla_{\mu}\phi\nabla_{\nu}\phi$.
The energy momentum tensor is
\ben
T_{\mu\nu}\equiv {2\over \sqrt {-g}}{\delta S_{k}\over \delta g^{\mu\nu}}= L_{X}\nabla_{\mu}\phi\nabla_{\nu}\phi - g_{\mu\nu}L
\label{eq:2}
\een
$L_{\mathrm X}= {dL\over dX},~~ L_{\mathrm XX}= {d^{2}L\over dX^{2}},
~~L_{\mathrm\phi}={dL\over d\phi}$ and  
$\nabla_{\mu}$ is the covariant derivative defined with respect to the gravitational metric $g_{\mu\nu}$.
The equation of motion is
\ben
-{1\over \sqrt {-g}}{\delta S_{k}\over \delta \phi}= \tilde G^{\mu\nu}\nabla_{\mu}\nabla_{\nu}\phi +2XL_{X\phi}-L_{\phi}=0
\label{eq:3}
\een
where  
\ben
\tilde G^{\mu\nu}\equiv L_{X} g^{\mu\nu} + L_{XX} \nabla ^{\mu}\phi\nabla^{\nu}\phi
\label{eq:4}
\een
and $1+ {2X  L_{XX}\over L_{X}} > 0$.

Carrying out the conformal transformation
$G^{\mu\nu}\equiv {c_{s}\over L_{x}^{2}}\tilde G^{\mu\nu}$, with
$c_s^{2}(X,\phi)\equiv{(1+2X{L_{XX}\over L_{X}})^{-1}}\equiv sound ~ speed $.

Then the inverse metric of $G^{\mu\nu}$ is   
\ben G_{\mu\nu}={L_{X}\over c_{s}}[g_{\mu\nu}-{c_{s}^{2}}{L_{XX}\over L_{X}}\nabla_{\mu}\phi\nabla_{\nu}\phi] 
\label{eq:5}
\een
A further conformal transformation \cite{gm1,gm2} $\bar G_{\mu\nu}\equiv {c_{s}\over L_{X}}G_{\mu\nu}$ gives
\ben \bar G_{\mu\nu}
={g_{\mu\nu}-{{L_{XX}}\over {L_{X}+2XL_{XX}}}\nabla_{\mu}\phi\nabla_{\nu}\phi}
\label{eq:6}
\een	
Here one must always have $L_{X}\neq 0$ for the sound speed 
$c_{s}^{2}$ to be positive definite and only 
then equations $(1)-(4)$ will be physically meaningful, 
since $L_{X}=0$ implies $L$ is independent of  
$X$, then from equation (\ref{eq:1}),  $L(X,\phi)\equiv L(\phi)$ i.e., $L$ becomes a function of pure potential and 
the very definition of $k-$essence fields becomes meaningless because such fields correspond to lagrangians where the kinetic energy dominates over the potential energy. Also the very concept of minimal coupling of $\phi$ to $g_{\mu\nu}$ becomes redundant, so the equation (\ref{eq:1}) meaningless and equations (\ref{eq:4}-\ref{eq:6}) ambiguous.

For the non-trivial configurations of the $k-$ essence field $\phi$, $\partial_{\mu}\phi\neq 0$ (for a scalar field,$\nabla_{\mu}\phi\equiv \partial_{\mu}\phi$ ) and  $\bar G_{\mu\nu}$ is not conformally 
equivalent to $g_{\mu\nu}$. So this $k-$ essence field $\phi$ field has the properties different from canonical scalar fields defined with $g_{\mu\nu}$ and the local causal structure is also different from those defined with $g_{\mu\nu}$.
Further, if $L$ is not an explicit function of $\phi$
then the equation of motion $(3)$ is reduces to;
\ben
-{1\over \sqrt {-g}}{\delta S_{k}\over \delta \phi}
= \bar G^{\mu\nu}\nabla_{\mu}\nabla_{\nu}\phi=0
\label{eq:7}
\een
We shall take the Lagrangian as $L=L(X)=1-V\sqrt{1-2X}$ with $V$ is a constant. 
This is a particular case of the DBI lagrangian \cite{gm1,gm2,born}
\ben
L(X,\phi)= 1-V(\phi)\sqrt{1-2X}
\label{eq:8}
\een
for $V(\phi)=V=constant$~~and~~$kinetic ~ energy ~ of~\phi>>V$ i.e.$(\dot\phi)^{2}>>V$. This is typical for the $k-$essence field where the kinetic energy dominates over the potential energy.
Then $c_{s}^{2}(X,\phi)=1-2X$.
For scalar fields $\nabla_{\mu}\phi=\partial_{\mu}\phi$. Thus (\ref{eq:6}) becomes
\ben
\bar G_{\mu\nu}= g_{\mu\nu} - \partial _{\mu}\phi\partial_{\nu}\phi
\label{eq:9}
\een
Note the rationale of using two conformal transformations:
the first is used to identify the inverse metric $G_{\mu\nu}$, while 
the second realises the mapping onto the   
metric given in $(9)$ for the lagrangian $L(X)=1 -V\sqrt{1-2X}$.

\section{Kerr-Newman metric and emergent gravity}
We consider the gravitational metric $g_{\mu\nu}$ is Kerr-Newman (KN) and denote
 $\partial_{0}\phi\equiv\dot\phi$, $\partial_{r}\phi\equiv\phi '$. We consider 
that the $k-$ essence scalar field $\phi\equiv\phi (r,t)$. The line element
of Kerr-Newman metric is \cite{kn}
\ben
ds^2_{KN}=f(r,\theta)dt^2-\frac{dr^2}{g(r,\theta)}+2H(r,\theta)d\phi dt
\nonumber\\-K(r,\theta)d\phi^2-\Sigma(r,\theta) d\theta^2
\label{eq:10}
\een
where,
$f(r,\theta)=\frac{\Delta(r)-\alpha^{2}sin^{2}\theta}{\Sigma(r,\theta)}$;\\
$g(r,\theta)=\frac{\Delta(r)}{\Sigma(r,\theta)}$;\\
$H(r,\theta)=\frac{\alpha sin^{2}\theta(r^{2}+\alpha^{2}-\Delta(r))}{\Sigma(r,\theta)}$;\\
$K(r,\theta)=\frac{(r^{2}+\alpha^{2})^{2}-\Delta(r)\alpha^{2} sin^{2}\theta}{\Sigma(r,\theta)}sin^{2}\theta$;\\
$\Sigma(r,\theta)=r^{2}+\alpha^{2}cos^{2}\theta$;\\
$\Delta(r)=r^{2}+\alpha^{2}+Q^2-2GMr$.
\vspace{0.2in}

It is to be noted that the above metric (\ref{eq:10}) also rediscovered in \cite{umetsu}. In \cite{mann}, elaborately shown how 
the Hawking temperature is not depending on $\theta$ although the metric functions depend on $\theta$. In our case the emergent gravity metric (\ref{eq:9}) $\bar G_{\mu\nu}$ contains extra terms (first derivative of $k-$essence scalar fields) {\it but these extra terms are still not depended on $\theta$. Therefore, the modified Hawking temperature will still be independent of $\theta$. 
For this reason, we will choose our evaluation for some fixed $\theta$, i.e., $\theta=0$ only.}
Assuming the Kerr-Newman metric along $\theta=0$. Then the above line element (\ref{eq:10}) becomes
\ben
ds^2_{KN,\theta=0}=F(r)dt^2-\frac{1}{F(r)}dr^2
\label{eq:11}
\een
with $\it{F(r)=\frac{\Delta(r)}{\Sigma}}$ and $\Sigma=r^2+\a^2$.

Also in \cite{umetsu1} was shown that the four dimensional spherically non-symmetric Kerr-Newman metric (\ref{eq:10}) transformed into a two dimensional spherically symmetric metric (\ref{eq:11}) in the region near the horizon by the method of dimensional reduction.

The emergent gravity metric (\ref{eq:9}) components are 
\ben
\bar G_{00}=g_{00}-(\partial _{0}\phi)^{2}={\Delta \over{\Sigma}}- \dot\phi ^{2}\nonumber\\
\bar G_{11}= g_{11} - (\partial _{r}\phi)^{2}= -{\Sigma\over{\Delta}} - (\phi ') ^{2}\nonumber\\
\bar G_{01}=\bar G_{10}=-\dot\phi\phi '. 
\label{eq:12}
\een

Then the emergent gravity line element (\ref{eq:12}) along $\theta=0$ becomes
\ben
ds^{2,emer}_{KN}=({\Delta \over{\Sigma}}- \dot\phi ^{2})dt^{2}
-({\Sigma\over{\Delta}} + (\phi ') ^{2})dr^{2}-2\dot\phi\phi 'dtdr
\nonumber\\
\label{eq:13}
\een
Now transform the coordinates \cite{gm1,gm2} from ($t,r$) to ($\omega,r$) such that
\ben
d\omega=dt-({\dot\phi \phi ' \over{{\Delta \over{\Sigma}}- \dot\phi^{2}}})dr
\label{eq:14}
\een
and considering
\ben
\dot\phi^{2}={\Delta^{2}\over{\S^{2}}}(\phi ')^{2}
\label{eq:15}
\een
we get the line element (\ref{eq:13}):

\ben
ds^{2,emer}_{KN}=({\Delta \over{\Sigma}}-\dot \phi^{2})d\omega^{2}
-\frac{dr^2}{({\Delta \over{\Sigma}}-\dot \phi^{2})}
\label{eq:16}
\een

We consider the solution of equation (\ref{eq:15}) as $\phi(r,t)=\phi_{1}(r)+\phi_{2}(t)$.

Then the equation (\ref{eq:15}) reduces to 
\ben
\dot\phi_{2}^{2}={\Delta^{2}\over{\S^2}}(\phi_{1} ')^{2}=K
\label{eq:17}
\een
where $K$ is a constant and $K\neq 0$ since $k-$essence scalar field will have {\it non-zero} kinetic energy. 
Now from (\ref{eq:17}) we get
$\dot \phi_{2}=\sqrt{K}$ and 
$\phi_{1} '=\sqrt{K}[\frac{(r^{2}+\alpha^{2})}{r^{2}-2GMr+\alpha^{2}+Q^2}]$

Therefore the solution of (\ref{eq:15}) is 
\ben
\phi(r,t)=\phi_{1}(r)+\phi_{2}(t)\nonumber\\
=\sqrt{K}[(r-GM)+GM~ln[(r-GM)^2\nonumber\\+\a^2+Q^2-(GM)^2]+\frac{2G^2 M^2-Q^2}{\sqrt{\a^2+Q^2-(GM)^2}}\nonumber\\tan^{-1}(\frac{r-GM}{\sqrt{\a^2+Q^2-(GM)^2}})] + \sqrt{k}t\nonumber\\
\label{eq:18}
\een
where
$\phi_{1}(r)=\sqrt{K}[(r-GM)+GM~ln[(r-GM)^2+\a^2+Q^2-(GM)^2]+\frac{2G^2 M^2-Q^2}{\sqrt{\a^2+Q^2-(GM)^2}}tan^{-1}(\frac{r-GM}{\sqrt{\a^2+Q^2-(GM)^2}})] $
and $\phi_{2}(t)=\sqrt{k}t$ and choosing integration constant to be zero.
Therefore the line element (\ref{eq:16}) becomes
\ben
ds^{2,emer}_{KN}=({\Delta \over{\S}}-K)d\omega^{2}-{1\over{({\Delta \over{\S}}-K)}}dr^{2}\nonumber\\
=\frac{\b\D'}{\S}d\omega^2-\frac{\S}{\b\D'}dr^2
\label{eq:19}
\een
where $\b=1-K$, $M'=\frac{M}{1-K}$, $\D'=r^2-2GM'r+Q'^2+\a^2$ and $Q'=\frac{Q}{\sqrt{1-K}}$.

This new metric (\ref{eq:19}) is also Kerr-Newman (KN) type along $\theta=0$ in the presence of dark energy.
{\it Note that $K\neq 1$ since $\beta$ cannot be zero, as then the metric (\ref{eq:19}) becomes singular. Also we have the total energy density is unity ($\Omega_{matter} +\Omega_{radiation} +\Omega_{dark energy}= 1$) \cite{gm2,wein}. So we can say that the dark energy density i.e., kinetic energy ($\dot\phi_{2}^{2}=K$) of $k-$essence scalar field (in unit of critical density) cannot be greater than unity.
Again $K$ cannot be greater than $1$ because the metric (\ref{eq:19}) will lead to wrong signature. The possibility of non-zero $K$ appears because that would imply the absence of dark energy. 
Therefore, the only allowed values of $K$ are  $0 < K < 1$. So there is no question of $K$ approaching towards unity and confusions regarding this limit is avoided}.
It can be shown that, for $r\rightarrow\infty$, this metric (\ref{eq:19}) is an approximate solution of Einstein's equation. 

Also mention that the mass and charge of this type black hole are modified as $M'=\frac{M}{1-K}$, $Q'=\frac{Q}{\sqrt{1-K}}$ respectively in the presence of dark energy density term $K=\dot\phi_{2}^2$.

Now we can show that the $k-$essence scalar field $\phi(r,t)$ given by equation (\ref{eq:18}) to satisfy the emergent equation of motion (\ref{eq:7}) along the symmetry axis $\theta=0$ at $r\rightarrow\infty$. For $\theta=0$, the emergent equation of motion (\ref{eq:7}) takes the form
\ben
\bar G^{00}\partial_{0}^{2}\phi_{\mathrm 2} 
+ \bar G^{11}\partial_{1}^{2}\phi_{\mathrm 1} 
-\bar G^{11}\Gamma_{11}^{1}\partial_{1}\phi_{\mathrm 1}\nonumber\\
+\bar G^{01}\nabla_{0}\nabla_{1}\phi
+\bar G^{10}\nabla_{1}\nabla_{0}\phi= 0.
\label{eq:20}
\een
The first term vanishes since $\phi_{2}(t)$ is linear in $t$ 
and the last two terms vanish because $\bar G^{01}=\bar G^{10}=0$. 

Using the expression for $$\Gamma_{11}^{1}= \frac{GM (\alpha^2-r^2)+Q^2 r}{(r^2+\alpha^2)(r^2-2GMr +\alpha^2+Q^2)}$$ 
the second and third terms for $r\rightarrow\infty$ goes as $\frac{(1-K)\sqrt{K}}{r^{2}}$. From the Planck collaboration results  \cite{planck1,planck2}, we have the value of dark energy density (in unit of critical density) $K$ is about $0.696$. Therefore, the second and third terms of (\ref{eq:20}) is negligible as the denominator goes to infinity. Therefore, in this limit the emergent equation of motion is satisfied. 

\section{The Hawking temperature for KN type metric in the presence of dark energy}

We use the tortoise coordinate defined by \cite{wheeler,umetsu1}
\ben
dr^{*}=\frac{dr}{f(r)}
\label{eq:21}
\een
with $f(r)=\frac{\b\D'}{\S}$ then the emergent line element (\ref{eq:19}) can be written as
\ben
ds^{2,emer}_{KN}=f(r)(d\o-dr^{*})(d\o+dr^{*})
\label{eq:22}
\een
At near the horizon the equation (\ref{eq:21}) can be written as
\ben
dr^{*}=\frac{(r^2+\a^2)dr}{\b(r-r_{+})(r-r_{-})}
\label{eq:23}
\een
with $r_{+}=GM'+\sqrt{(GM')^2-Q'^2-\a^2}$ and $r_{-}=GM'-\sqrt{(GM')^2-Q'^2-\a^2}$.
Integrating equation (\ref{eq:23}) we get
\ben
r^{*}= \frac{1}{\b}[r+({r_{+}^{2}+\alpha^{2}\over{r_{+}-r_{-}}})ln~|r-r_{+}|\nonumber\\
+({r_{-}^{2}+\alpha^{2}\over{r_{-}-r_{+}}})ln~|r-r_{-}|]+C
\label{eq:24}
\een
where $C$ is an integration constant.

The above equation (\ref{eq:24}) can be written in terms of surface gravity when $r>r_{+}$ as \cite{umetsu1}
\ben
r^{*}=\frac{r}{\b}+\frac{1}{2\chi_{+}}ln~(\frac{r-r_{+}}{r_{+}})+\frac{1}{2\chi_{-}}ln~(\frac{r-r_{-}}{r_{-}})
\label{eq:25}
\een
with surface gravity ($+$ sign for outer horizon and $-$ sign for inner horizon)
\ben
\chi_{\pm}\equiv\frac{1}{2}f'(r)\mid_{r=r_{\pm}}=\frac{\b}{2}[\frac{r_{\pm}-r_{\mp}}{r_{\pm}^{2}+\a^2}].
\label{eq:26}
\een

Also we calculate the Hawking temperature \cite{haw} for (\ref{eq:19}) using {\it tunneling formalism} \cite{mann,mitra,jiang,murata,ma} for the two horizons as follows.

We going over to the Eddington-Finkelstein coordinates $(v,r)$ or 
$(u,r)$ along $\theta=0$ i.e., {\it introducing advanced and retarded null coordinates} \cite{gm2} $$v=\omega+r^{*}~~;~~u=\omega-r^{*}$$. Using this coordinate the line element
(\ref{eq:19}) becomes
\ben
ds^{2,emer}_{KN}=({\beta\Delta' \over{r^{2}+\alpha^{2}}})dv^{2}-2dv dr\nonumber\\
={\beta(r-r_{+})(r-r_{-}) \over{r^{2}+\alpha^{2}}}dv^{2}-2dv dr.
\label{eq:27}
\een

Also we calculate the Hawking temperature \cite{haw} for (\ref{eq:27}) using {\it tunneling formalism} \cite{mann,mitra,jiang,murata,ma} for the two horizons as follows.

A massless particle in a black hole background is described by the Klein-Gordon equation
\ben
\hbar^2(-\bar G)^{-1/2}\partial_\mu( \bar G ^{\mu\nu}(-\bar G)^{1/2}\partial_\nu\Psi)=0.
\label{eq:28}
\een
We can expands $\Psi$ as
\ben
\Psi=exp({i\over{\hbar}}S+...)
\label{eq:29}
\een
to obtain the leading order in $\hbar$ the Hamilton-Jacobi equation is
\ben
\bar G^{\mu\nu}\partial_\mu S \partial_\nu S=0.
\label{eq:30}
\een
We consider $S$ is independent of $\theta$ and $\phi$. Then the above equation (\ref{eq:30})
\ben
2{\partial S\over{\partial v}}{\partial S\over{\partial r}}+
(\frac{\beta(r^{2}-2GM^{'}r+\alpha^{2}+Q'^{2})}{r^{2}+\alpha^{2}})({\partial S\over{\partial r}})^{2}=0\nonumber\\
\label{eq:31}
\een
The action $S$ is assumed to be of the form
\ben
S=-Ev+W(r)+J(x^{i})
\label{eq:32}
\een
Then 
\ben
\partial_{v}S=-E~;~\partial_{r}S=W^{'}~;~\partial_{i}S=J_{i}
\label{eq:33}
\een
$J_{i}$ are constants chosen to be zero.
Now putting the values of equation (\ref{eq:33}) in equation (\ref{eq:31}) we get
\ben
-2EW^{'}(r)+(\frac{\beta(r^{2}-2GM^{'}r+\alpha^{2}+Q'^{2})}{r^{2}+\alpha^{2}})
(W^{'}(r))^{2}=0.\nonumber\\
\label{eq:34}
\een
Then 
\ben
W(r)=\int \frac{[E(r^{2}+\alpha^{2})+E(r^{2}+\alpha^{2})]dr}{\beta(r-r_{+})(r-r_{-})}\nonumber\\
=2\pi i ({E\over{\beta}})({r_{+}^{2}+\alpha^{2}\over{r_{+}-r_{-}}})
+2\pi i ({E\over{\beta}})({r_{-}^{2}+\alpha^{2}\over{r_{-}-r_{+}}})\nonumber\\
=W(r_{+})+W(r_{-})
\label{eq:35}
\een
The two values of $W(r)$ correspond to the outer and inner horizons respectively.

Therefore the equation (\ref{eq:32}) becomes
\ben
S=-Ev+2\pi i ({E\over{\beta}})({r_{+}^{2}+\alpha^{2}\over{r_{+}-r_{-}}})
+2\pi i ({E\over{\beta}})({r_{-}^{2}+\alpha^{2}\over{r_{-}-r_{+}}})+J(x^{i})\nonumber\\
\label{eq:36}
\een
So the tunneling rates are
\ben
\Gamma^{KN}_{+emergent} \sim e^{-2~Im S{+}} \sim e^{-2~Im W(r_{+})}\nonumber\\
=e^{-4\pi ({E\over{\beta}})({r_{+}^{2}+\alpha^{2}\over{r_{+}-r_{-}}})}=e^{-{E\over{K_{B}T_{+}}}}
\label{eq:37}
\een
and
\ben
\Gamma^{KN}_{-emergent} \sim e^{-2~Im S{-}} \sim e^{-2~Im W(r_{-})}\nonumber\\
=e^{-4\pi ({E\over{\beta}})({r_{-}^{2}+\alpha^{2}\over{r_{-}-r_{+}}})}=e^{-{E\over{K_{B}T_{-}}}}
\label{eq:38}
\een
where $K_{B}$ is Boltzman constant.
From these above two expressions (\ref{eq:37}) and (\ref{eq:38}) the corresponding Hawking temperatures of the two horizons are 
\ben
T_{+emergent}^{KN}=\frac{\hbar c^{3}\beta}{4\pi k_{B}}(\frac{r_{+}-r_{-}}{r_{+}^{2}+\alpha^{2}})~~~~~~~~~~~~~~~~~~~~~~~~~~~~~~\nonumber\\
=\frac{\hbar c^{3}\beta}{2\pi k_{B}}[\frac{\sqrt{(GM')^2-\a^2-Q'^2}}{2GM'(GM'+\sqrt{(GM')^2-\a^2-Q'^2})-Q'^2}]\nonumber\\
\label{eq:39}
\een
and 
\ben
T_{-emergent}^{KN}=\frac{\hbar c^{3}\beta}{4\pi k_{B}}(\frac{r_{-}-r_{+}}{r_{-}^{2}+\alpha^{2}})~~~~~~~~~~~~~~~~~~~~~~~~~~~~~~~~~\nonumber\\
=-\frac{\hbar c^{3}\beta}{2\pi k_{B}}[\frac{\sqrt{(GM')^2-\a^2-Q'^2}}{2GM'(GM'-\sqrt{(GM')^2-\a^2-Q'^2})-Q'^2}]\nonumber\\
\label{eq:40}
\een
with $\beta=1-K$.

The usual Hawking temperature for Kerr-Newman black hole is \cite{mann}
\ben
T_{\pm}^{KN}=\frac{\hbar c^{3}}{4\pi k_{B}}(\frac{r_{\pm}-r_{\mp}}{r_{\pm}^{2}+\alpha^{2}})~~~~~~~~~~~~~~~~~~~~~~~~~~~~~~~\nonumber\\
=\frac{\hbar c^{3}}{2\pi k_{B}}[\frac{\sqrt{(GM)^2-\a^2-Q^2}}{2GM(GM\pm\sqrt{(GM)^2-\a^2-Q^2})-Q^2}]\nonumber\\
\label{eq:41}
\een
{\it The above temperatures (\ref{eq:39},\ref{eq:40}) are modified in the presence of dark energy. These temperatures are different from usual Hawking temperature (\ref{eq:41}) as the presence of terms $\b=1-K$, $M'=\frac{M}{1-K}$ and $Q'=\frac{Q}{\sqrt{1-K}}$ where $K$ is the dark energy density (in unit of critical density).}

\section{Kerr-Newman-AdS background}
We consider the gravitational metric $g_{\mu\nu}$ is Kerr-Newman-AdS (KNAdS). 
The line element of KNAdS metric \cite{jiang,murata,ma,cald,aliev} is
\ben
ds^2_{KNAdS}=\frac{1}{\S}[\D_{r}-\D_{\theta}\a^2 sin^{2}\theta]dt^2
-\frac{\S}{\D_{r}}dr^2-\frac{\S}{\D_{\theta}}d\theta^2\nonumber\\-\frac{1}{\S(\Xi)^2}[\D_{\theta}(r^2+\a^2)^2-\D_{r}\a^2 sin^{2}\theta]sin^{2}\theta~d\phi^2\nonumber\\+\frac{2\a}{\S\Xi}[\D_{\theta}(r^2+\a^2)-\D_{r}]sin^{2}\theta~dtd\phi ~~~~~~~~~
\label{eq:42}
\een
where
\ben
\S=r^2+\a^2 cos^{2}\theta ;~~~\Xi=1-\frac{\a^2}{l^2}
\label{eq:43}
\een
\ben
\D_{\theta}=1-\frac{\a^2}{l^2}cos^{2}\theta;~
\D_{r}=(r^2+\a^2)(1+\frac{r^2}{l^2})-2GMr+Q^2.\nonumber\\
\label{eq:44}
\een
The parameters $M$ and $\a$ are related to the mass and
angular momentum of the black hole, $G$ is the gravitational constant and $l$ is the curvature radius determined by the negative cosmological constant ($\L<0$) $\Lambda=-\frac{3}{l^2}$.

Again we choose symmetric axis along $\theta=0$ as before since in \cite{mann} elaborately shown that the Hawking temperature is independent of $\theta$.
Then the line element (\ref{eq:42}) reduces to
\ben
ds^2_{KNAdS,\theta=0}=F(r)dt^2-\frac{1}{F(r)}dr^2
\label{eq:45}
\een
with $F(r)=\frac{\D_{r}}{\S}$ and $\S=r^2+\a^2$.

Using this (\ref{eq:45}) the emergent gravity metric (\ref{eq:9}) components are
\ben
\bar G_{00}=g_{00}-(\partial _{0}\phi)^{2}={\D_{r} \over{\Sigma}}- \dot\phi ^{2}\nonumber\\
\bar G_{11}= g_{11} - (\partial _{r}\phi)^{2}= -{\Sigma\over{\D_{r}}} - (\phi ') ^{2}\nonumber\\
\bar G_{01}=\bar G_{10}=-\dot\phi\phi '. 
\label{eq:46}
\een
Again we consider the $k-$essence scalar field $\phi(r,t)$ is spherically symmetric. So the emergent gravity line element for KNAdS background along $\theta=0$ is
\ben
ds^{2,emer}_{KNAdS}=({\D_{r} \over{\Sigma}}- \dot\phi ^{2})dt^{2}
-({\Sigma\over{\D_{r}}} + (\phi ') ^{2})dr^{2}-2\dot\phi\phi 'dtdr.
\nonumber\\
\label{eq:47}
\een
Transform the coordinates $(t,r)$ to $(\o,r)$ as
\ben
d\omega=dt-({\dot\phi \phi ' \over{{\D_{r} \over{\Sigma}}- \dot\phi^{2}}})dr
\label{eq:48}
\een
and we choose 
\ben
\dot\phi^{2}={\D_{r}^{2}\over{\S^{2}}}(\phi ')^{2}.
\label{eq:49}
\een

Then the line element (\ref{eq:47}) becomes
\ben
ds^{2,emer}_{KNAdS}=({\D_{r} \over{\S}}-\dot \phi^{2})d\omega^{2}
-\frac{dr^2}{({\D_{r} \over{\S}}-\dot \phi^{2})}
\label{eq:50}
\een

We consider again the solution of equation (\ref{eq:49}) as $\phi(r,t)=\phi_{1}(r)+\phi_{2}(t)$.

Then the equation (\ref{eq:49}) is 
\ben
\dot\phi_{2}^{2}={\D_{r}^{2}\over{\S^2}}(\phi_{1} ')^{2}=K
\label{eq:51}
\een
where $K$ is a constant and $K\neq 0$.
From (\ref{eq:51}) we get
$\dot \phi_{2}=\sqrt{K}$ and 
$\phi_{1} '=\sqrt{K}[\frac{(r^{2}+\alpha^{2})}{(r^{2}+\a^2)(1+\frac{r^2}{l^2})-2GMr+Q^2}]$.
So the solution of equation (\ref{eq:49}) is
\ben
\phi(r,t)=\phi_{1}(r)+\phi_{2}(t)~~~~~~~~~~~~~~~~~~~~~~~~~~~~~~~~~~~~~~\nonumber\\
=\frac{C\sqrt{K}}{2}~ln|r^2+\l r+m|+\frac{D\sqrt{K}}{2}~ln|r^2-\l r+n|\nonumber\\+\frac{\sqrt{K}(2A-\l C)}{2\sqrt{m-\frac{\l^2}{4}}}~tan^{-1}(\frac{r+\frac{\l}{2}}{\sqrt{m-\frac{\l^2}{4}}})~~~~~~~~~~~~~~~~\nonumber\\
+\frac{\sqrt{K}(2B+\l D)}{2\sqrt{n-\frac{\l^2}{4}}}~tan^{-1}(\frac{r-\frac{\l}{2}}{\sqrt{n-\frac{\l^2}{4}}})
+\sqrt{K}t\nonumber\\
\label{eq:52}
\een
where 
\ben
\phi_{1}(r)=\s{K}\int{\frac{(r^{2}+\alpha^{2})}{(r^{2}+\a^2)(1+\frac{r^2}{l^2})-2GMr+Q^2}dr}\nonumber\\
=\s{K}\int{\frac{(r^{2}+\alpha^{2})}{(r^2+\l r+m)(r^2-\l r+n)}dr}\nonumber\\
=\frac{C\sqrt{K}}{2}~ln|r^2+\l r+m|+\frac{D\sqrt{K}}{2}~ln|r^2-\l r+n|\nonumber\\+\frac{\sqrt{K}(2A-\l C)}{2\sqrt{m-\frac{\l^2}{4}}}~tan^{-1}(\frac{r+\frac{\l}{2}}{\sqrt{m-\frac{\l^2}{4}}})~~~~~\nonumber\\
+\frac{\sqrt{K}(2B+\l D)}{2\sqrt{n-\frac{\l^2}{4}}}~tan^{-1}(\frac{r-\frac{\l}{2}}{\sqrt{n-\frac{\l^2}{4}}})~~~~~~
\label{eq:53}
\een
and 
\ben
\phi_{2}(t)=\s{K}t.
\label{eq:54}
\een
Now we clarify the parameters of the above equation (\ref{eq:52}):
$C=\frac{-1}{2\l^2+m-n}~,~D=\frac{1}{2\l^2+m-n}~,~$
$A=\frac{1}{m+n}[\a^2-\frac{n(m-n)}{2\l^2+m-n}]~,~$ 
$B=\frac{1}{m+n}[\a^2+\frac{m(m-n)}{2\l^2+m-n}]~,~$ 
$\l=[(\frac{1}{2}(-T+\s{T^2+4H^3}))^{1/3}+(\frac{1}{2}(-T-\s{T^2+4H^3}))^{1/3}-\frac{2(l^2+\a^2)}{3}]^{1/2}~,~$
$m=\frac{1}{2}[(l^2+\a^2)+\l^2+\frac{2GMl^2}{\l}]~,~$
$n=\frac{1}{2}[(l^2+\a^2)+\l^2-\frac{2GMl^2}{\l}]~,~$
$H=-\frac{1}{9}[l^4+2(3\a^2+2Q^2)l^2+\a^4]~,~$
$T=-\frac{1}{27}[2l^6-6(12Q^2+11\a^2-18G^{2}M^2)l^4-6\a^2(12Q^2+11\a^2)l^2+2\a^6]$.

For this type of $k-$essence scalar field $\phi$ (\ref{eq:52}), the line element (\ref{eq:50}) reduces to
\ben
ds^{2,emer}_{KNAdS}=({\D_{r} \over{\S}}-K)d\omega^{2}-{1\over{({\D_{r} \over{\S}}-K)}}dr^{2}\nonumber\\
=\frac{\b\D_{r}'}{\S}d\omega^2-\frac{\S}{\b\D_{r}'}dr^2
\label{eq:55}
\een
where $\b=1-K$, $M'=\frac{M}{1-K}$, $\D_{r}'=(r^2+\a^2)(1+\frac{r^2}{l'^2})-2GM'r+Q'^2$, $Q'=\frac{Q}{\sqrt{1-K}}$ and $l'^2=(1-K)l^2$.
Similar reasons as before here also the only allowed values of $K$ are  $0 < K < 1$. Also it can be shown that this metric (\ref{eq:55}) is an approximate solution of Einstein's equations at $r\rightarrow\infty$ along $\theta=0$. Note that the parameters $M, Q, l$ are also modified in the presence of dark energy density ($K$).

We can show that the $k-$essence scalar field (\ref{eq:52}) is satisfied emergent gravity equation of motions (\ref{eq:7}) along $\theta=0$ at $r\rightarrow\infty$. For $\theta=0$, the emergent equation of motion (\ref{eq:7}) takes the form
$\bar G^{00}\partial_{0}^{2}\phi_{\mathrm 2} 
+ [\bar G^{11}\partial_{1}^{2}\phi_{\mathrm 1} 
-\bar G^{11}\Gamma_{11}^{1}\partial_{1}\phi_{\mathrm 1}]
+\bar G^{01}\nabla_{0}\nabla_{1}\phi
+\bar G^{10}\nabla_{1}\nabla_{0}\phi= 0.$
The first term vanishes since $\phi_{2}(t)$ is linear in $t$ 
and the last two terms vanish because $\bar G^{01}=\bar G^{10}=0$. 
Using the value of $$\Gamma_{11}^{1}=\frac{1}{\S\D}[\frac{-2r^5}{l^2}+\frac{2\pi \a^2 r^3}{3}-GM(r^2-\a^2)+(Q^2-\frac{\a^4}{l^2})r]$$ we get 
the terms within third bracket are vanished at  $r\rightarrow\infty$.

\section{The Hawking temperature for KNAdS type metric in the presence of dark energy}
We calculate the Hawking temperature using tunneling formalism \cite{mitra,ma,cald,aliev}. The horizons of the metric (\ref{eq:55}) in the presence of dark energy are determined by
\ben
\D_{r}'=(r^2+\a^2)(1+\frac{r^2}{l'^2})-2GM'r+Q'^2\nonumber\\
=\frac{\b}{l'^{2}}[r^4+r^2(\a^2+l'^2)-2GM'l'^{2}r+l'^{2}(\a^2+Q'^2)]\nonumber\\
=\frac{\b}{l'^{2}}(r-r_{++}^{d})(r-r_{--}^{d})(r-r_{+}^{d})(r-r_{-}^{d})=0~~~~~~
\label{eq:56}
\een
The equation $\D_{r}'=0$ has four roots, two real positive roots and two complex roots. We denote $r_{++}^{d}$ and $r_{--}^{d}$ are complex roots and $r_{+}^{d}$ and $r_{-}^{d}$ are positive real roots in the presence of dark energy $(K)$. Here we consider $r_{+}^{d}>r_{-}^{d}$ so that $r_{+}^{d}$ is the black hole event horizon and $r_{-}^{d}$ is the Cauchy horizon of the KNAdS type black hole (\ref{eq:55}).

Now we use the Eddington-Finkelstein coordinates $(v,r)$ or 
$(u,r)$ along $\theta=0$ i.e., {\it advanced and retarded null coordinates} \cite{gm2} $$v=\omega+r^{*}~~;~~u=\omega-r^{*}$$ with 
\ben
dr^{*}=\frac{(r^2+\a^2)dr}{\frac{\b}{l'^{2}}(r-r_{++}^{d})(r-r_{--}^{d})(r-r_{+}^{d})(r-r_{-}^{d})}
\label{eq:57}
\een
we get the emergent gravity line element (\ref{eq:55}) becomes
\ben
ds^{2,emer}_{KNAdS}=\frac{\b\D_{r}'}{\S}dv^{2}-2dvdr\nonumber\\
=[\frac{\frac{\b}{l'^{2}}(r-r_{++}^{d})(r-r_{--}^{d})(r-r_{+}^{d})(r-r_{-}^{d})}{r^2+\a^2}]dv^2-2dvdr.\nonumber\\
\label{eq:58}
\een
Proceedings exactly same as KN type case we can calculate the Hawking temperatures for KNAdS type black hole (\ref{eq:58}) as:
\ben
T^{KNAdS}_{+emergent}=\frac{\hbar c^{3}\beta}{4\pi k_{B}l^2}[\frac{(r_{+}^{d}-r_{++}^{d})(r_{+}^{d}-r_{--}^{d})(r_{+}^{d}-r_{-}^{d})}{(r_{+}^{d})^2+\a^2}]\nonumber\\
=-\frac{\hbar c^{3}(1-K)\L}{12\pi k_{B}}[\frac{(r_{+}^{d}-r_{++}^{d})(r_{+}^{d}-r_{--}^{d})(r_{+}^{d}-r_{-}^{d})}{(r_{+}^{d})^2+\a^2}]~~~~~~~~
\label{eq:59}
\een
and

\ben
T^{KNAdS}_{-emergent}=\frac{\hbar c^{3}\beta}{4\pi k_{B}l^2}[\frac{(r_{-}^{d}-r_{++}^{d})(r_{-}^{d}-r_{--}^{d})(r_{-}^{d}-r_{+}^{d})}{(r_{-}^{d})^2+\a^2}]\nonumber\\
=-\frac{\hbar c^{3}(1-K)\L}{12\pi k_{B}}[\frac{(r_{-}^{d}-r_{++}^{d})(r_{-}^{d}-r_{--}^{d})(r_{-}^{d}-r_{+}^{d})}{(r_{-}^{d})^2+\a^2}]~~~~~~~~
\label{eq:60}
\een
where $k_{B}$ is the Boltzman constant. These temperatures $T^{KNAdS}_{+emergent}$ and $T^{KNAdS}_{-emergent}$ are different from usual Hawking temperature for KNAdS black hole as reported on \cite{jiang}-\cite{aliev}.
Here $\L<0$, $r_{+}^{d}$ and $r_{-}^{d}$ are positive and $r_{+}^{d}>r_{-}^{d}$;  $r_{++}^{d}$ and $r_{--}^{d}$ are complex conjugate, these make sure that the temperature of event horizon is positive.

Note that an Anti-de Sitter (AdS) space has negative cosmological constant in a vacuum, where empty space itself has negative energy density but positive pressure, unlike our accelerated Universe where observations of distant supernovae indicate a positive cosmological constant corresponding to the de-Sitter space \cite{wein} which has positive energy density but negative pressure. Dark energy is one of the candidate being regarded as the origin of this accelerated expansion where pressure is negative. So for AdS space, the cosmological constant is negative which cannot be associated with dark energy. Therefore, here the  dark energy comes only from the $k-$essence scalar field. Also note that the Hawking temperature for KNAdS black hole is already eavaluated in \cite{ma}.

\section{Conclusion}
In this work we have determined the Hawking temperatures in the presence of dark energy for emergent gravity metrics having 
Kerr-Newman and Kerr-Newman-AdS backgrounds.
We have shown that in the presence of dark energy the Hawking temperatures are modified. We did the calculation 
for Kerr-Newman and Kerr-Newman-AdS background metrics along $\theta=0$ since the Hawking temperature is independent of $\theta$ and show that the modified metrics i.e., emergent gravity black hole metrics for both cases are satisfies Einstein's equations for large $r$ and the emergent black hole always radiates. These new emergent gravity metrics are mapped on to the Kerr-Newman and Kerr-Newman-AdS type metrics.
Throughout the paper our analysis is done in the 
context of dark energy in an emergent gravity scenario having $k-$essence scalar fields $\phi$ with a Dirac-Born-Infeld type 
lagrangian. In both cases the 
scalar field $\phi(r,t)=\phi_{1}(r)+\phi_{2}(t)$ also satisfies the emergent gravity equations of motion at  
$r\rightarrow\infty$ for $\theta=0$. 

\vspace{0.3in}
~~~~~~~~~~~~~~~~~~~~~~~~~~~~~~***~~~~~~~~~~~~~~~~~~~~~~

The authors would like to thank the referees for illuminating suggestions to improve the manuscript.

\end{document}